\documentclass[11pt,titlepage]{article}

\usepackage{amsmath}
\usepackage{amssymb}

\usepackage{graphicx} 

\usepackage{color}         

\usepackage{amsmath} 
\usepackage{amssymb} 

\newcommand{\expe}[1]{e^{#1}}
\newcommand{\abs}[1]{\left | #1 \right |}

\newcommand{\vect}[1]{\ensuremath{\mathbf{#1}}}

\newcommand{\vchem}{\ensuremath{\mathbf{v_\chi}}}

\newcommand{\divg}{\nabla\cdot}
\newcommand{\Lap}{\nabla^2}

\newcommand{\fig}[1]{Fig. \ref{#1}}  

\newcommand{\appdx}[1]{Appendix \ref{#1}} 

\newcommand\be{\begin{equation}}
\newcommand\ee{\end{equation}}
\newcommand\bea{\begin{eqnarray}}
\newcommand\eea{\end{eqnarray}}

\newcommand{\eqn}[1]{equation (\ref{#1})}  
\newcommand{\eq}[1]{(\ref{#1})}            

\newcommand{\ut}[1]{\,\rm{#1}}  

\newcommand{\colisp}{{\it E. coli }} 

\usepackage[round,numbers,sort&compress]{natbib} 
\title{Migration of chemotactic bacteria in soft agar: role of gel concentration}

\author{Ottavio A. Croze\thanks{
           Corresponding author.  Email: o.croze@physics.org} \\
        School of Mathematics and Statistics\\
            University of Glasgow\\
                Glasgow G12 8QW, UK.
    \and Gail P. Ferguson \\
    School of Medicine and Dentistry\\
    Division of Applied Medicine\\
            Institute of Medical Sciences\\
                University of Aberdeen \\
                    Foresterhill, Aberdeen AB25 2ZD, UK.
    \and Michael E. Cates \\
        SUPA, School of Physics and Astronomy  \\
            University of Edinburgh \\
                Mayfield Road, Edinburgh EH9 3JZ, UK.
    \and Wilson C. K. Poon \\
        SUPA, School of Physics and Astronomy  \\
            University of Edinburgh \\
                Mayfield Road, Edinburgh EH9 3JZ, UK.}

\date{}

\begin{document}

\maketitle

\abstract{We study the migration of chemotactic wild type {\it Escherichia coli} populations in semi-solid (`soft') agar in the concentration range $C=0.15-0.5\%$(w/v). For $C \lesssim 0.35\%$, expanding bacterial colonies display characteristic chemotactic `rings'. At $C = 0.35\%$, however, bacteria migrate as broad circular bands rather than sharp rings. These are growth/diffusion waves arising because of suppression of chemotaxis by the agar and have not been previously reported experimentally. For $C=0.4-0.5\%$ expanding colonies do not span the depth of the agar and develop pronounced front instabilities. The migration front speed is weakly dependent of agar concentration below $C = 0.25\%$, but decreases sharply above this value. We discuss these observations in terms of an extended Keller-Segel model for which we derived novel transport parameter expressions accounting for perturbations of the chemotactic response by collisions with the agar. The model allows to fit the observed front speed decay in the range $C=0.15-0.35\%$, and its solutions qualitatively reproduce the observed transition from chemotactic to growth/diffusion bands. We discuss the implications of our results for the study of bacteria in porous media and for the design of improved bacteriological chemotaxis assays.\\

}

\clearpage

\section*{Introduction}

Much is understood about the motility of flagellated bacteria in open liquid media \citep{Berg04} and on solid surfaces \citep{Harshey03}. In contrast, bacterial locomotion within semi-solid media is much less well studied, even though bacteria often colonise three dimensional semi-solid environments, e.g. host tissues or foodstuffs. `Soft' agar with concentration $0.2\% \lesssim C \lesssim 0.35\%$ (throughout, \% = \%w/v), a gel network whose main component is the semi-flexible polysaccharide agarose \citep{Rochas06}, is a plausible model for many of these kinds of solid environments. It has been used in studies aimed at understanding motile microbial pathogens growing inside the semi-solid matrix of a variety of foods and of infected hosts, and therefore for predicting spoilage and infection \citep{Wimpenny95,Wimpenny97, VibrioInfect91}.

Soft agar was first introduced into microbiology for assaying chemotaxis \citep{Adler66}. Chemotactic wild-type {\em Escherichia coli} inoculated at one end of a capillary filled with nutrient buffer spread out in `bands' as they successively deplete the medium of oxygen and various nutrients. In a Petri dish filled with nutrient agar, the observation of successive sharp circular bands (`rings') progressing outwards from the colony inoculated into the centre of the soft agar is taken to confirm the chemotaxis genotype, since non-chemotactic mutants spread out uniformly \citep{WolfeBerg89}.

Interestingly, the agar concentration in this widely-used chemotaxis assay is not standardized, varying from investigator to investigator, or even within the same study, in the range of $0.2\% \lesssim C \lesssim 0.4\%$. The tacit assumption seems to be that, as long as concentrations are in the `soft range', agar conveniently suppresses thermal and biological convection in the liquid medium which hosts bacteria, but otherwise has no interesting effect. Wolfe and Berg's investigation  \citep{WolfeBerg89} of a number chemotactic mutants in soft agar ($C = 0.2\% - 0.35\%$) appears to confirm this assumption. They report no qualitative difference except a slowing down of the spreading front. 

However, the run-and-tumble motion of {\it E. coli} \cite{Berg04} and similar bacteria suggests {\it a priori} that soft agar should affect chemotaxis. The pore size of soft agar is $\sim 1 \mu$m \citep{Righetti81}, the same order as typical bacterial run lengths. Since cells perform chemotaxis by altering their tumble frequency and therefore run length, the structure of agar may therefore interfere with the ability to chemotactically `bias' random walks. The expectation is that this interference will be more pronounced in agar than in other porous media, such as sand and soil, where most pore sizes ($\approx 10 \mu\mbox{m}$ to 1~mm) are very much larger than typical run lengths. Indeed, a study of {\it Pseudomonas putida} in sand columns with grain sizes in the range $80-800\mu$m found no effect on chemotaxis \cite{BartonFord95}.

In this work, we show by experiment and theory that the chemotaxis of {\it E. coli} in soft agar is indeed strongly affected as the concentration is increased in the range $C = 0.15- 0.5\%$. We observe that, as $C$ is increased, the colony growth pattern changes qualitatively. Also, the speed of the migrating front is weakly dependent on $C$ at $C \lesssim 0.25\%$, but decreases sharply with agar concentration above this value. To understand our observations, we formulate a modified Keller-Segel type model \citep{KellerSegel72, LKA84} with transport parameters that are functions of agar concentration. These functions are derived by extending a recent description of bacterial chemotactic response \citep{DeGennes04}, and account for the diminished ability of bacteria to detect chemical gradients in dense gels. Our model is able to fit quantitatively the observed dependence of front speed on agar concentration. 


\section*{Methods}

We used the {\it E. coli} K-12 derivative AB1157, a chemotactic wild-type \citep{DeWittAdelberg62}. 
Plate cultures from frozen stocks were refrigerated at 4$^{\circ}$C for up to 3 weeks prior to use. Luria Broth (LB) agar was prepared by adding 1.5-5 g/l of Difco Bacto-agar to LB Broth (10 g/l Difco Bacto-Tryptone, 5 g/l Difco Yeast extract, 10 g/l NaCl) \citep{Sambrook89}. The mixture was autoclaved at 121$^{\circ}$C under 1.02~atm for 30~min, and left to cool for an hour. Agar plates were then prepared by pouring hot (45-50$^{\circ}$C) nutrient agar into standard sized (100~mm diameter, 12~mm deep) plastic Petri dishes (Sterilin) on a level surface; 58~ml were poured in each plate (final agar thickness: $10.0 \pm 0.1$~mm). Poured plates were left to set for about a day at ambient temperature (22-27$^{\circ}$C). The final pH of the agar was $7.3 \pm 0.1$.

Late exponential phase cultures (OD$_{600} = 0.8-1.3$) were prepared by inoculating single colonies in LB-filled flasks incubated at $30.0^{\circ}$C and shaken at 200~rpm. Cultures were then diluted to OD$_{600} = 0.1$ ($\approx 10^8$ viable cells/ml) and agar plates were inoculated by delivering a droplet of $\lesssim 1\mu$l via a $2\mu$l pipette. Inocula were left to sit on the agar for about an hour; then a thin layer (thickness $\approx 1.5$~mm, corresponding to $\approx 9$~ml) of sterile filtered mineral oil (Sigma) was poured on the plates. Except at the highest agar concentrations studied ($C \gtrsim 0.4\%$), the inoculum had been (to the naked eye) completely assimilated by the time the oil was poured, with no visible evidence of the pouring spreading organisms to other parts of the agar surface. At the highest $C$, the inoculum was not significantly incorporated into the agar after an hour, so that careful pouring was needed to minimise spreading. When significant spreading did occur, the plates were discarded. The thin oil layer kept evaporative losses to $< 1\%$ (weight) during our observations, but did not generate anaerobic conditions \citep{Rahn41}.

The Petri dishes were incubated at $30.0 \pm 0.5 ^{\circ}$C on a dark background, illuminated from the sides, and imaged at 30 minute intervals using computer-controlled CCD cameras. Images were analysed using ImageJ (NIH) and IDL  (RSI, Boulder, CO): a, static noise (by subtracting the first frame in each sequence) was removed; b, slightly non-uniform illumination was corrected for (subtracting the image background fitted using a sliding paraboloid with 20 pixels rolling ball radius); c, histograms were matched histograms using gray-scale mapping \citep{BenkeHedger96}. Then we obtained azimuthally-averaged radial intensity profiles from those images without significant blebs (see below for observations and further discussion on blebs). The images presented in \fig{015-04radprof}a and \fig{MorphoHighC} were processed using only steps b and c.

At high $C$, image thresholding enabled us to determine the colony area, $A$, from which we calculated the radius $r = \sqrt{A/\pi}$. At lower $C$ we fitted a circle to the intensity maximum in each image and determined the area of the fitted circle, from which $r$ was then calculated. 

\section*{Model of chemotactic \colisp populations in agar} 

The fundamental processes in agar plates inoculated with bacteria are growth due to nutrient uptake and dispersion due to chemotactic motility, which can be modelled by generalized Keller-Segel models \citep{KellerSegel72, MainiArmitage08}. Migrating populations of bacteria in agar have been described for bacteria chemotactically sensing nutrients \citep{NossalTh72, Russians93} or attractants secreted by the cells themselves \citep{MurrayBudreneBerg95}. However, these models ignore, or inconsistently account for, the effect of agar on bacterial chemotaxis. Since agar is a porous gel, one might think that existing descriptions of other porous media \citep{BartonFord96, HarveyFord07} should be applicable to agar. We will show below that these descriptions are incorrect. The model presented here is an adaptation of that originally formulated by Lauffenburger, Kennedy and Aris (LKA) for a chemotactic population with growth in one dimension \citep{LKA84}. Our model differs from the LKA model in three ways. First, we model growth as logistic, while LKA used a linear term. Secondly, we work in two dimensions, since LKA's 1D analysis is not adequate for modelling Petri dishes for early times. Finally and most crucially, bacterial transport coefficients in our model are not constants, but functions of agar concentration derived from a recent model of chemotactic response \cite{DeGennes04}.

\subsection*{Model equations}

The starting point equations of our model are: 
\bea
&&\frac{\partial b}{\partial t}=-\divg\left[-\vect{\mu}(s, C)\nabla b+ \vchem(s, C) b\right]+k_g b \left(g(s) - \frac{b}{k_b}\right) \label{BacFlux1}\\
&&\frac{\partial s}{\partial t}=D\Lap s -\frac{1}{Y} k_g g(s) b \label{SubFlux1}.
\eea
Equation \ref{BacFlux1} expresses the conservation of bacteria, with population density $b(\vect{r},t)$. This population evolves in response to the combined effect of its diffusive and chemotactic fluxes, with diffusivity $\mu(s, C)$ and drift velocity $\vchem(s, C)$; these are in general functions of both substrate and agar concentrations, $s(\vect{r},t)$ and $C$ respectively. The bacterial population also evolves by growth, with birth rate $k_g g(s)$, where $k_g$ is the maximum growth rate and  $g(s)$ is a function of substrate uptake, and a death rate $-k_g b/k_b$, where $k_b$ is the carrying capacity of the population. Equation (\ref{SubFlux1}) models the conservation of the first, most readily metabolised substrate, with concentration $s(\vect{r},t)$, diffusing with diffusivity $D$ and being consumed by bacteria at a rate $-k_g g(s)$. In tryptone broth or LB this substrate is L-serine \citep{Adler66, Russians93}. In the consumption term of \eq{SubFlux1}, $Y$  is the bacterial yield upon consumption ($b=Y s$). We now assume the following relations: $\mu(s, C)= \mu(C)$; $\vchem(s, C)= \chi(C)\,\nabla f_\chi(s)$, where $f_\chi(s)= \frac{s}{s+k_\chi}$; and $g(s)=\frac{s}{s+k_s}$. That is, we assume the diffusivity to be isotropic and independent of substrate concentration; $\mu$ depends only on the spatially uniform agar concentration $C$ (see below). The chemotactic velocity, $\vchem$, is assumed linear in the gradient of a `receptor-adsorption' function $f_\chi(s)$; $k_{\chi}$ is the characteristic saturation concentration of the chemotactic response \citep{LapidusSchiller76}. The proportionality constant is the chemotactic coefficient $\chi$ which, like $\mu$, is assumed to depend only on agar concentration. The relations involving $\vchem$ and $\mu$ are approximate forms valid in the limit of shallow concentration gradients \citep{ChenFordCummingsLim98}. Bacterial growth depends on substrate concentration through a Monod-type growth function $g(s)$ \citep{Monod49}; $k_s$ is the characteristic saturation concentration for growth.

We have derived the dependence of the diffusivity, $\mu$, and chemotactic parameter, $\chi$, on agar concentration by modifying de Gennes' integral model of bacterial chemotactic response \citep{DeGennes04}, as detailed in \appdx{agarappendix1}. The model quantifies the intuition that in a dense matrix of obstacles bacteria are reoriented by collisions with the matrix as well as by tumbles, making chemotaxis inefficient. Such collisions increase the effective bacterial tumble rate from the {\it in liquido} value $\alpha_0$ to $\alpha(C)=\alpha_0(1+f(C)$, where the function $f(C)$ quantifies the collision-induced concentration-dependent increase of the tumble rate (see below).  From our model it can be shown (see \appdx{agarappendix1}) that the chemotactic transport parameters in agar are given by: 
\be\label{TranspMicroCexp0}
\mu(C)=\mu_0 \left[1+f(C)\right]^{-1};\,\,\,\,\,\,\,\,  
\chi(C)= \chi_0 \left[1+f(C)\right]^{-2}I_\chi \left[f(C)\right]
\ee
where $\mu_0$ and $\chi_0$ are the bacterial diffusivity and chemotactic coefficient in the absence of agar. The agar concentration-dependent integral $I_\chi$ is given by \eqn{chemointegral} and its value depends on the form of the chemotactic response function, $K(t)$, see \eqn{chemoresponsefunction}. The function $f(C)$ gauges the increase with agar concentration of the tumble rate {\it in agar}, $\alpha_A(C)$, with respect to its {\it in liquido} value $\alpha_0$. That is, we assume $\alpha_A(C)=\alpha_0 f(C)$. Since collisions are more frequent for a higher density of obstacles, we expect $f(C)$ to monotonically increase with $C$. We adopt the ansatz $f(C)=\exp[\frac{C-C_1}{C_0}]$, where $C_0$ is a characteristic concentration. The concentration `shift' $C_1$ accounts for the possibility that the tumble rate in agar can recover its {\it in liquido} value for small, but nonzero agar concentrations: $\alpha_A(C\le C_1)\to 0$ so $\alpha(C\le C_1)\to\alpha_0$, $C_1>0$. 

To further understand the experimentally observed migration transition and to compare our results to those derived for bacteria in porous media \citep{BartonFord96, HarveyFord07}, we have also derived asymptotic limits to the expression \eq{TranspMicroCexp0} for $\chi(C)$. As shown in \appdx{agarappendix1}, these asymptotic limits are: 
\bea\label{TranspMicro-Lim1-0}
\chi(C)\simeq
\begin{cases} 
\chi_0 \left[1+f(C)\right]^{-2}\left[1- \kappa f(C)\right] & \text{if $\tilde{\alpha}(C)\approx 1$}\\\label{TranspMicro-vLim1}
\beta\, \chi_0 \left[1+f(C)\right]^{-3}&\text{if $\tilde{\alpha}(C)\gg 1$}.
\end{cases}
\eea
where $\kappa$ and $\beta$ are constants (see \appdx{agarappendix1}) and $\tilde{\alpha}(C)=1+\alpha_A/\alpha_0=1+f(C)$ is the dimensionless effective tumble rate in agar (see \eq{agaralpha}). The limits \eq{TranspMicro-Lim1-0} reflect the effect of confinement in agar on chemotaxis. At low concentrations, agar does not significantly impede chemotaxis, and bacteria can tumble relatively freely: $\tilde{\alpha}(C)\approx 1$ (`efficient' limit). At higher concentrations, frequent collisions with the agar $\tilde{\alpha}(C)\gg 1$ confuse the chemotactic response (the `confused limit'). Neither expressions \eq{TranspMicroCexp0}, nor the limits \eq{TranspMicro-Lim1-0} coincide with those derived in previous models of bacteria in porous media. These models treat bacteria in porous media like molecular gases, and so derive transport parameters obeying the balance relation: $\frac{\mu(C)}{\mu_0}=\frac{\chi(C)}{\chi_0}=[1+\frac{\alpha_A(C)}{\alpha_0}]^{-1}$ \citep{BartonFord96, HarveyFord07}. However, in our model this relation does not hold (even for very low $C$) because of the effect of collisions on chemotaxis. It is only recovered in the absence of agar. We assume changes in the swimming speed, $D$ or $k_g$ with agar concentration are negligible \citep{SchantzLauffer62, SharmaKnapp93}.

\subsection*{Geometry, scaling and model parameters}

We will consider only the two-dimensional, axisymmetric limit of our equations. Our assumption is that, modulo a time shift, fully developed bacterial front dynamics are insensitive to particular initial conditions (provided the initial colony is azimuthally symmetric). The characteristic length and time scales for our experiments are millimeters and hours, so we rescale our equations by $\tau_g\sim k_g^{-1}$, where $k_g$ is the growth rate, and $l_g\sim\sqrt{\mu_0/k_g}$, the average length a cell diffuses during a doubling time (in the absence of agar).  The population density is rescaled by its carrying capacity, $k_b$, and all concentrations by the initial substrate concentration, $s_0$. We rescale all diffusivities by that of the bacterial population (in the absence of agar), $\mu_0$. We also rescale the yield $Y$ by the maximum possible yield $k_b/s_0$. Thus:
$r= r \sqrt{k_g/\mu_0};\,
T= k_g t;\,
B= b/k_b;\, 
S= s/s_0;\,
N= D/\mu_0;\,
\delta_0= \chi_0/\mu_0;\,
K_\chi= k_\chi/s_0;\,
K_s= k_s/s_0;\,
H= k_b/(Y\,s_0);\,
M(C)=\mu(C)/\mu_0;\,
X(C)=\chi(C)/\chi_0$.
The model equations (\ref{BacFlux1}) and (\ref{SubFlux1}) in dimensionless form then read
\bea
&\frac{\partial B}{\partial T}= M(C) \Lap B - \delta_0 X(C) \divg \left(B\,\frac{d F_\chi}{d S}\nabla S \right) + B \left[G(S) - B\right] \label{BacFlux1DL},\\
&\frac{\partial S}{\partial T}=N \Lap S -H \,G(S)\, B \label{SubFlux1DL},
\eea
where $F_\chi(S)= \frac{S}{S+K_\chi}, G(S)=\frac{S}{S+K_s}$ and where the dependence on agar concentration is through the functions:
\be\label{TranspMicroCexp0dimless}
M(C)=\left[1+f(C)\right]^{-1};\,\,\,\,\,\,\,\,  
X(C)= \left[1+f(C)\right]^{-2}I_\chi \left[f(C)\right],
\ee
where we recall $f(C)= \exp[\frac{C-C_1}{C_0}]$,  where $C_0$ and $C_1$ are the characteristic concentrations introduced earlier. The parameter $\delta_0$ is significant: it measures the relative magnitude of chemotactic advection to random diffusion in the absence of agar (a chemotactic P\'{e}clet number). It is how this ratio is modified by agar which leads to surprising results, as we will see. We have also defined the dimensionless parameters $N$, the ratio of nutrient and bacterial diffusivities, and $H$, the ratio of carrying capacitance of the bacterial population to the maximum population obtainable from the nutrient available. Equations \eq{BacFlux1DL} and \eq{SubFlux1DL} are subject to no-flux boundary conditions and to the initial conditions:
\bea
&&B(R,0)=\expe{-\frac{R^2}{\sigma^2}};\mbox{~~}S(R,0)=1-\expe{-\frac{R^2}{\sigma^2}}, \label{IC}
\eea
where $\sigma$ is the width of an initial Gaussian packet of bacteria. The following parameter values were used to solve our equations: growth rate, $k_g=0.7$h$^{-1}$ (from {\it in liquido} growth curve); initial cell concentration, $b_0(=k_b)=3.5\times10^8 \ut{cells/ml}$ (from viable counts); initial substrate 
concentration (of L-serine in LB), $s_0=1\ut{mM}$ ($5$-$8\ut{mM}$ \citep{LBCompositionPaper}); cell diffusivity (no agar), 
$\mu_0=5.7\ut{mm}^2/h$ ($1.2\ut{mm}^2/h$ \citep{AhmedStocker08}); chemotactic parameter (no agar), 
$\chi_0=600\ut{mm}^2/h$ ($450\ut{mm}^2/h$ to $\alpha$-methylaspartate \citep{AhmedStocker08}); substrate diffusivity, $D=3\ut{mm}^2/h$; chemotactic threshold concentration, $k_{\chi}=0.5\ut{mM}$ ($0.2\ut{mM}$ for $\alpha$-methylaspartate \citep{AhmedStocker08}); growth threshold concentration, $k_s=1\ut{mM}$; yield, $Y=10^{11}\ut{cells/ml/M}$. The growth rate and initial cell concentration were determined by our own independent experiments indicated in brackets. All other parameters are based on experimental literature values for {\it E. coli}, many of which have been used in other models of {\it E. coli} migration \cite{MurrayII, LapidusSchiller78}. Reference literature values close to parameters we changed are reported in brackets above. In addition to these macroscopic parameters, we use the {\it in liquido} tumble rate $\alpha_0=1$ s and the constant $A_0=0.5$ to calculate the integral  $I_\chi$ in \eq{TranspMicroCexp0dimless} using \eq{chemointegral} and \eq{chemoresponsefunction} \citep{ClarkGrant05}. The concentrations $C_0$ and $C_1$ are free parameters, fixed by fitting the predicted front speeds with those we observed experimentally (see Results and Discussion). Prior to performing the fit, the values of the parameters $k_g$, $\mu_0$, $\chi_0$ and $k_\chi$ were adjusted slightly to match the values of experimental and predicted band speed for $C=0.15\%$ (assuming this is the same as {\it in liquido}).

With these parameters, the dimensionless constants of the model have the values:
$\delta_0=105;\,
K_\chi= 0.53;\,
K_s= 1;\,
N= 0.5;\,
H= 3.5$.
The above parameters will not be changed in our investigation and the initial packet width $\sigma$ is fixed at $2$. Equations \eq{BacFlux1DL} and \eq{SubFlux1DL} in 1D axisymmetric form were solved numerically for $C=0.15-0.35\%$ on a linear domain ($L=100$) using Matlab subject to initial conditions \eq{IC} and no-flux boundary conditions. Migration front speeds were obtained by subtracting the position of the leading edge inflexion points of solution profiles calculated at neighbouring time points and dividing by the time interval. Like in experiment, these speeds were calculated in the linear growth regime (long times) where speed does not change with radius. 

\section*{Results}

\subsection*{Observations on migration morphology and radial dynamics}

We first report qualitative features of colony morphology and dynamics. For all concentrations studied ($C = 0.15-0.5\%$) it takes 5-7 hours for the bacterial inoculum on the agar surface to become visible. The inoculum then grows in optical density and, after an additional time lag of 1-50 hours (likely caused by the oil overlay, but with no influence on the reproducibility of subsequent front dynamics), the initial bacterial colony migrates across the plates. Stills from early and advanced stages of colony migration for concentrations in the range $0.15\% \leq C \leq 0.35\%$, are shown in \fig{015-04radprof}a. Two striking effects of increasing concentration are immediately apparent: the change from a morphology displaying sharp rings to one which is more diffuse and featureless, and the loss of circular symmetry in the advanced stage of migration at high concentrations ($C \approx 0.35\%$).

At the lowest concentrations sampled, $C = 0.15 - 0.2\%$, bacteria migrate as sharp circular bands inside the agar. We observed two bands in succession. The first band sharpens as it migrates across the plate, \fig{015-04radprof}b,c. The second band is slower than the first and also appears to sharpen as it travels. Interestingly, the first band at $C = 0.15\%$ initially displays internal structure (a double band, see first frame of \fig{015-04radprof}b) and is reflected from the plate walls (not shown) before the second band catches up with it. 

Bacteria also migrate as circular bands for $C = 0.25 - 0.3\%$. Again two bands were observed, but now they travelled together \fig{015-04radprof}d,e. At $C = 0.35\%$, sharp bands are no longer visible (\fig{015-04radprof}f). The colony grows from the inoculum as a circular disk with a slightly nebulous front (\fig{015-04radprof}). The intensity across the disk is initially approximately uniform, falling off at the edges, defining the band front \fig{015-04radprof}f. At later times, however, it displays a broad band structure. We did not follow the radial development of these bands to the edge of the plate because the colony front develops instabilities (blebs) that disrupt circular symmetry.  

Visual inspection confirmed that bacteria had spread from the surface inoculum into the agar to a significant depth. For concentrations supporting bands these are initially hard to resolve for radii smaller than the agar depth: the colony appears like a uniform expanding circle from above (\fig{015-04radprof}a, top row). For larger radii the first band is visible and clearly spans the depth of the agar, as observed by Adler \citep{Adler66}. For colonies with two distinct bands, it is not clear at what depths the second bands occur; from our images they seem to be further inside the agar. Microscopy (not shown) reveals that for $C<0.4\%$ bacteria penetrate significantly beyond $1$mm in depth, but for this and larger concentrations it seems that agar limits penetration to a few mm from the surface.

\begin{figure}[tbph]
\begin{center}  
\includegraphics[width=3.25in]{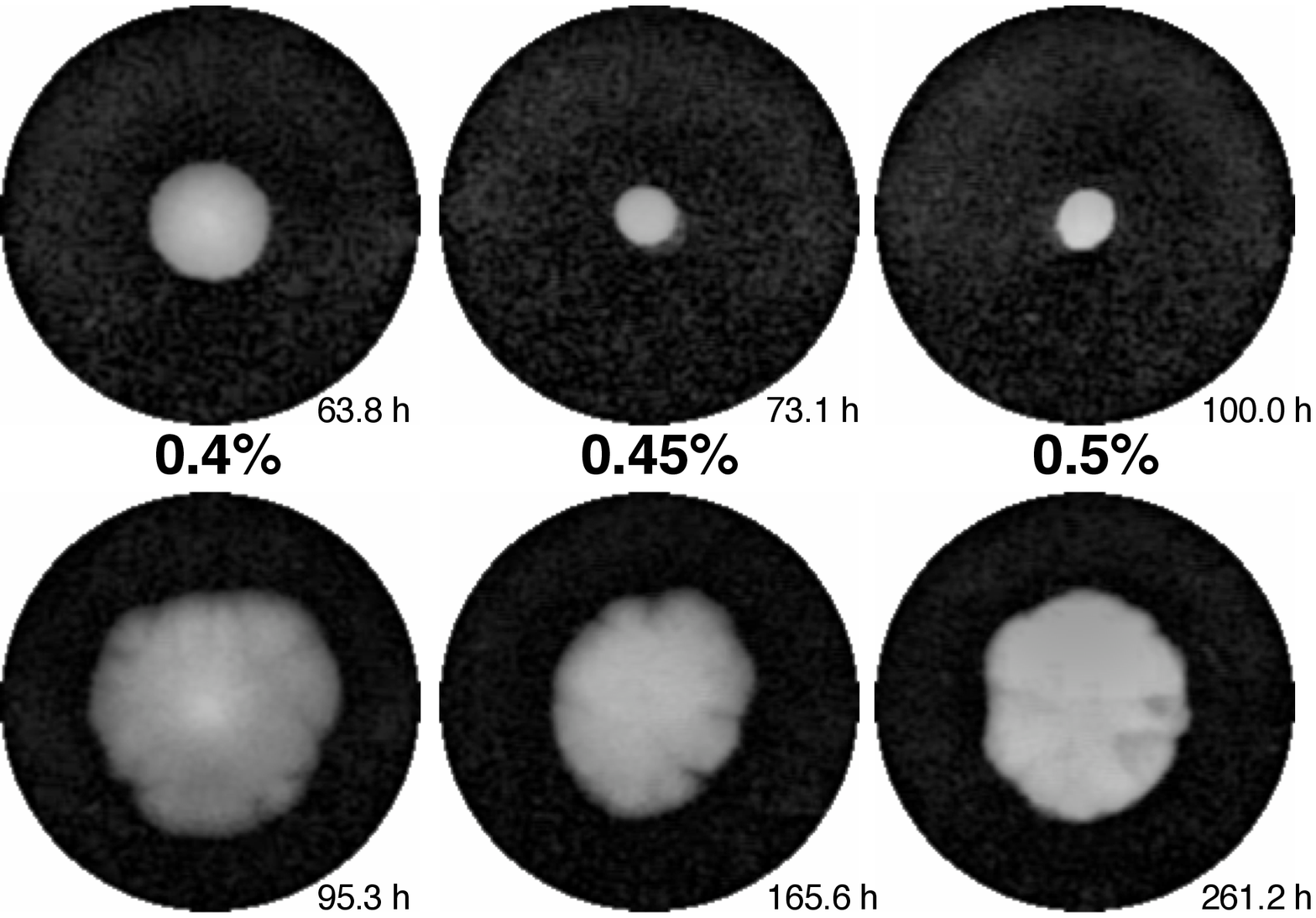} 
\caption{}
\label{MorphoHighC} 
\end{center}
\end{figure}

At $C = 0.4 - 0.5\%$, shown in \fig{MorphoHighC}, the expanding colony appeared as homogeneous, solid circles initially. However, the front invariably developed extensive blebs. The blebbing instability set in earlier for higher concentrations (e.g. at $C = 0.4\%$ blebs appeared when the colony radius was beyond a third of the plate radius, while at $C = 0.45\%$ it appeared at around one sixth). The blebs developed into wedge shaped sectors, giving the colony an overall flower shape (\fig{MorphoHighC}). At these concentrations, the colony also appeared to spread on the surface of the agar (though not by classical `swarming'), but we did not investigate such surface migration. 

\subsection*{Effect of concentration on radial migration}

\begin{figure}[tbph]
\begin{center}  
\includegraphics[width=3.25in]{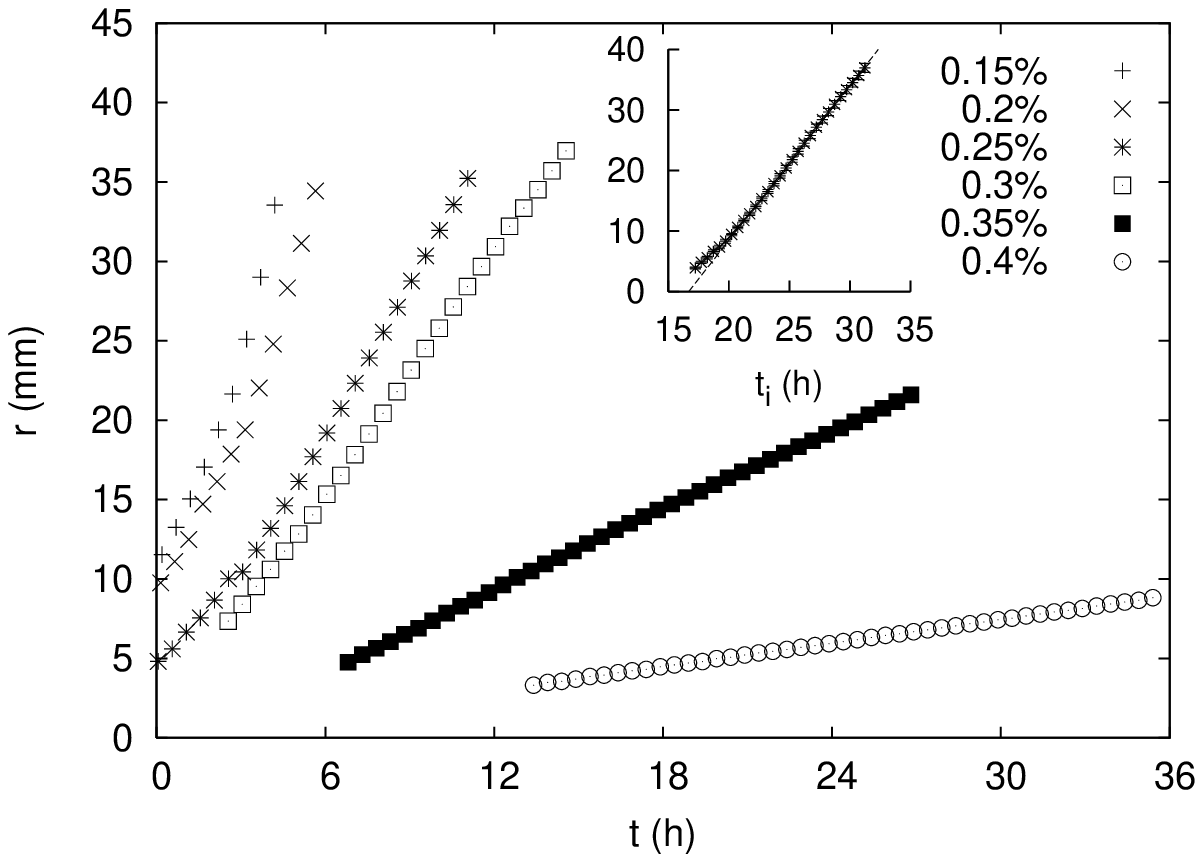} 
\caption{}
\label{Rvst015-04} 
\end{center}
\end{figure}

In \fig{Rvst015-04} we plot the radius of the outermost migrating front (band), $r$, against the time,  $t=t_i-\Delta t_{l}$, elapsed since inception of visible colony growth, where $t_i$ is the time since inoculation and $\Delta t_{l}$ is the latency time before a colony grows out. We estimated $\Delta t_{l}$ from the intersection of a linear fit to the raw radial data with the time axis (\fig{Rvst015-04}, inset). A substantial portion of the radial growth is linear in time for $0.15\% < C < 0.4\%$. Linear portions can also be identified for $0.45$ and $0.5\%$ (not shown), though the extent of these data is severely limited by the formation of blebs. Slopes from the fits to the radial growth curves in the range $0.15-0.35\%$ ($0.15-0.5\%$) are plotted as a function of agar concentration in \fig{VvsCa} (and inset). At $C \leq 0.25\%$ the migration speed is at best weakly affected by concentration, but beyond this value it decreases dramatically. Our model can account for this behaviour (see below).

\subsection*{Theoretical front speed decay and band profiles}

\begin{figure}[tbph]
\begin{center}  
\includegraphics[width=3.25in]{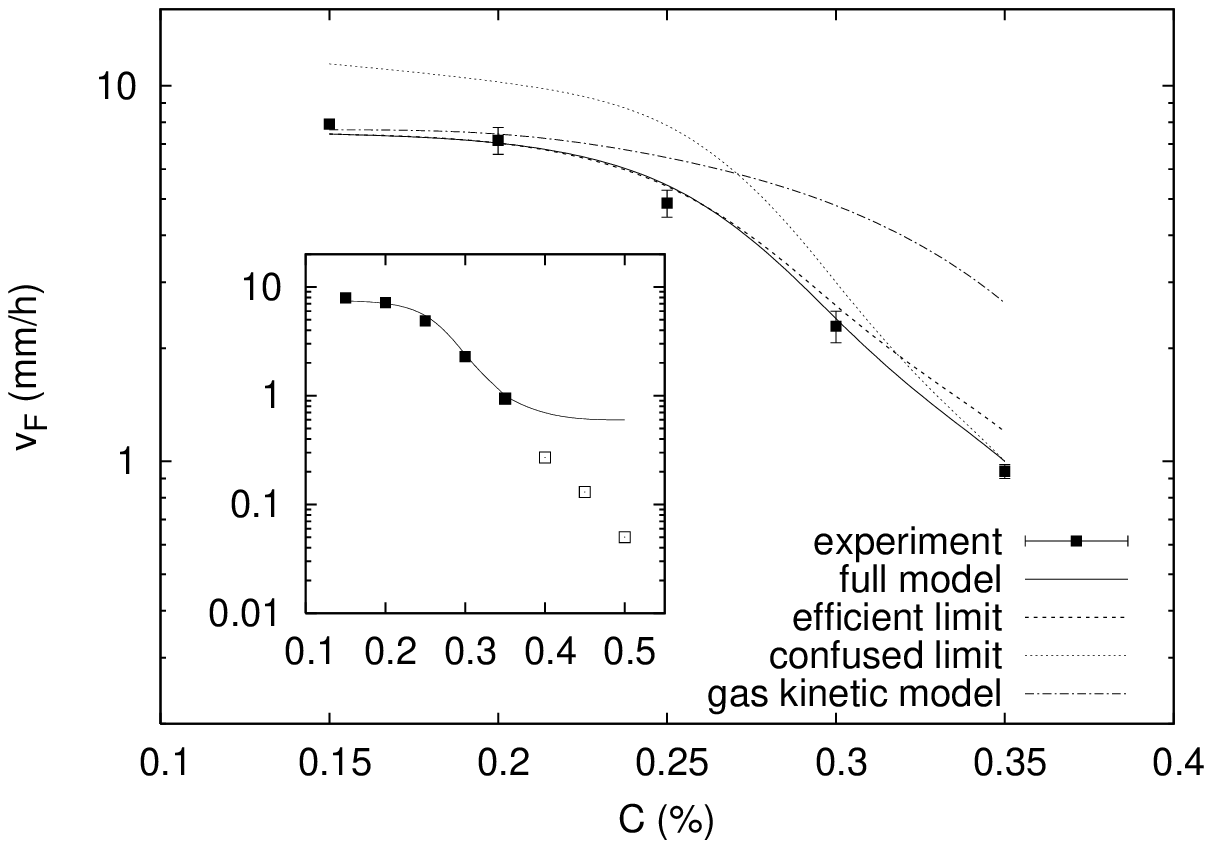} 
\caption{}
\label{VvsCa} 
\end{center}
\end{figure}

A fit to the experimental front speed data from solutions of our full model using relations \eq{TranspMicroCexp0} is shown in \fig{VvsCa}; also shown are the `efficient' and `confused' limits of the model for the same parameters. As expected, the efficient (confused) limit is a good description at low (high) concentration. The evolution of the theoretical band profiles corresponding to the full model best fit is shown in \fig{thprofiles} (left). Also shown in \fig{thprofiles} (right) is the prediction using transport parameters from gas kinetic models derived for bacteria in porous media: $\frac{\mu(C)}{\mu_0}=\frac{\chi(C)}{\chi_0}=[1+\frac{\alpha_A(C)}{\alpha_0}]^{-1}$ \citep{BartonFord96, HarveyFord07}. As concentration is increased in the experimental range $C = 0.15-0.35\%$, the full model band profiles displays a gradual transition from sharp, chemotaxis-dominated bands to broader, growth/diffusion-dominated bands. In the gas kinetic model, because the chemotaxis parameter, $\chi$, and diffusivity, $\mu$, have the same functional dependence on $C$, profiles remain sharp for all concentrations. The rounded profiles predicted by our model arise from suppression (`confusion') of the chemotactic response caused by bacterial collisions with the agar. When the chemotactic flux in equation \eq{BacFlux1} becomes negligible with respect to the fluxes due to logistic growth and diffusion, the travelling band solutions to \eq{BacFlux1} and \eq{SubFlux1} change from sharp, fast chemotaxis-dominated bands to slower, broader bands driven by growth/diffusion processes. This is what we observe experimentally. The breakdown of the model in the range $C = 0.4$-$0.5\%$, evident from the inset to \fig{VvsCa}, is explained in the discussion below.

\section*{Discussion}

We have experimentally studied the migration of chemotactic \colisp populations in soft agar of concentration in the range $C=0.15-0.5\%$. Consistently with other investigators we find that increasing agar concentration decreases the speed of propagation of the bacterial front \citep{WolfeBerg89, Wimpenny97, Eihaetal02} and severely hampers penetration for $C\gtrsim 0.5\%$ \citep{Wimpenny97, Eihaetal02}. However, our work also reveals a hitherto unobserved transition in the dynamics of the population as agar concentration increases. The gradual transition is from a dynamics displaying characteristic sharp chemotactic bands (rings) to one where the bacteria travel as broader bands. By increasing the chemotaxis to diffusion ratio $\delta_0=\chi_0/\mu_0$ Lauffenburger {\it et al.} (LKA) theoretically studied the transition from sharp chemotactic to broader growth/diffusion bands, but failed to find evidence for the latter in studies of chemotaxis in capillaries \citep{LKA84}. Interestingly, we have discovered that sufficiently concentrated agar provides an environment where chemotaxis is suppressed and growth/diffusion processes can be observed to dominate the band dynamics.

To understand our experimental results we also built a model of bacterial migration in agar. We extended the LKA model and coupled it to the first full expressions for the concentration dependence of bacterial diffusivity $\mu(C)$ and chemotactic parameter $\chi(C)$ in agar. We derived these (see \appdx{agarappendix1}) from an adaptation to agar of de Gennes' model of bacterial chemotactic response  \citep{DeGennes04}. Collisions with the matrix of concentrated agar (effective tumble rate $\alpha(C)=\alpha_0(1+f(C))$, where $f(C)= \exp[\frac{C-C_1}{C_0}]$) confuse this response causing $\mu(C)$ and $\chi(C)$ to have different functional forms. Our model can thus predict the band transition we observe experimentally. We obtained a best fit of the theoretical front speeds to the experimentally observed values (\fig{VvsCa}) in the concentration range $0.15$-$0.35\%$, finding the characteristic concentrations $C_0=0.035\%$ and $C_1=0.28\%$. In comparing model profiles, \fig{thprofiles}, with experimental ones, \fig{015-04radprof}b-f, we note that the vertical axes in the latter probably do not map linearly to cell density due to multiple scattering effects. In addition, `dead' or non-motile bacteria contribute to the signal but do not contribute in theoretical plots. With these caveats, we see that for $C=0.15$-$0.35\%$ our model qualitatively reproduces the experimentally observed transition in the colony (band) profile at long times rather nicely (\fig{thprofiles}, left): bands change from sharp to broad as concentration in increased. 

The model breaks down for $C=0.4$-$0.5\%$. At these concentrations bacterial diffusivity becomes very small (e.g. $M(0.4\%)=0.03$) and equations \eq{BacFlux1} and \eq{SubFlux1} predict a front speed independent of $C$. However, the measured (early, bleb-less) front speed continues to fall sharply with $C$, see inset to \fig{VvsCa}. One reason the model fails is that small diffusivity affects growth at high $C$. During a doubling time bacteria in $0.4\%$ agar are able to diffuse $\approx 6$ times less far than for $C=0.15\%$ ({\it in liquido}), which increases competition for nutrients with neighbours. Further, at high $C$ small bacterial diffusivity means growth is limited by that of nutrients: $\mu(0.4\%)/D=0.06$. Diffusion limited growth is known to produce branching instabilities like those we observe \citep{BenJacob00}.To fully explain high concentration colony morphologies (\fig{MorphoHighC}) changes in gene expression in response to low nutrient levels will also need to be considered. An interesting possibility is that in high $C$ agar cell densities could reach large enough values to elicit quorum sensing responses \citep{SuretteBassler99}. Experimentally, the situation for $C\geq0.4\%$ is also complicated by the observation of coexisting subpopulations (see results and also \citep{Eihaetal02}), one growing on the surface and one in the bulk, which does not penetrate very deeply (the dynamics is no longer 2D as assumed). Modelling these very different conditions is left to a future study.
 
We have so far been implicitly discussing the first (front) band. Experimentally a second band is also observed for $C<0.35\%$ which, as agar concentration is increased, travels closer and closer to the first (see \fig{015-04radprof}b-f). As mentioned bacteria in LB preferentially metabolise one nutrient at a time: the first band aerobically consumes L-serine and the second L-aspartate, with a roughly constant metabolic delay $T_m$ between bands \citep{Adler66, Russians93, WolfeBerg89}. Thus the maximum spacing between bands $L_b\sim v_F(C) T_m$ will decay with $C$ like $v_F(C)$, the speed of the first band. In this paper the emphasis has been on explaining the experimentally observed shape transition of the first band. In the future, it will be interesting to extend our model and experiments to quantify chemotaxis and its suppression for all nutrients consumed. Accounting for multiple bands, as well as using improved receptor-adsorption functions for growth and chemotaxis, will allow more realistic predictions for the trailing edge of the bands.

The suppression of chemotaxis we have studied is relevant to the migration of bacteria in porous materials other than agar, important in bioremediation \citep{HarveyFord07} and food spoilage \citep{Wimpenny97}. As discussed above, previous gas kinetic models of bacterial migration in porous media  \citep{BartonFord96, HarveyFord07} do not account for the possibility of the chemotactic response becoming confused by collisions with agar. This neglect, which is an implicit consequence of assuming bacterial populations behave like molecular gases, invalidates the predictions of these models in porous media with a finite concentration of obstacles, even if dilute. Gas kinetic models can provide good fits to our experimental front speed data (with different values for the characteristic concentrations $C_0$ and $C_1$), but cannot also reproduce the experimentally observed transition in front shapes. On the other hand, provided pores are larger than a cell, our model accurately describes the transport of chemotactic bacteria in general porous media. 

Our results also have potentially important implications for microbiological practice. Microbiologists studying motility often make chemotactic mutants which are screened for using chemotaxis assays. One of these assays, the `motility assay', involves inoculating soft agar and imaging the resulting bacterial colony, like we have done in this study. The agar concentration for such assays is not standard (values in the range $0.1$-$0.4\%$ can be found in the literature \citep{BarakEisenbach99, Proteus0p4agar00}), and seems to be a matter of convenience (e.g. larger concentration allows to study more than one colony in the same plate \citep{HPylMotAss10}). When chemotactic mutants are screened for, the chemotactic band phenotype is sought for as a marker of chemotaxis, its absence denoting a successful chemotactic mutant \citep{WolfeBerg89} or a failed restoration of the chemotaxis phenotype \citep{BarakEisenbach99}. Our experiments suggest, however, that chemotactic run-and-tumble bacteria above a certain (still soft) concentration of agar will fail to show the band phenotype. Thus, if agar plates are used to assay for chemotaxis it will be important take into account the possibility that suppression of the band phenotype by the physical environment may occur. Performing assays at a number of agar concentrations spanning the ÔsoftÕ range (0.1-0.4\%) should therefore be part of standard protocol when screening for chemotaxis. 

\section*{Acknowledgements}

We acknowledge work by Jessica Cameron in the embryonic stages of this research, assistance by Sarah Spragg, and discussions with Rosalind Allen, Julien Tailleur, Davide Marenduzzo and {\it with Gary Dorken}. OAC and WCKP were funded by the EPSRC EP/E030173 and EP/D071070. GPF was funded by an MRC New Investigator grant G0501107. MEC was funded by the Royal Society.


\begin{appendix}
\section{Modelling run-and-tumble chemotaxis in agar}\label{agarappendix1}
\renewcommand{\thesection}{\Alph{section}}
\setcounter{equation}{0}
\renewcommand{\theequation}{\thesection\arabic{equation}}

Using a microscopic model of run-and-tumble dynamics in one dimension (see \citep{TailleurCates08} and references therein) it can be shown that the bacterial diffusivity, $\mu$, and the chemotactic parameter, $\chi$, are given by:
\be\label{TranspMicro}
\mu=\frac{2 v^2}{\alpha^{+}+\alpha^{-}};\,\,\,\,  
v_\chi=v\frac{\alpha^{-} - \alpha^{+}}{\alpha^++\alpha^-}
\ee
where $\alpha^{\pm}$ are the mean tumble probabilities for bacteria moving up ($+$) and down ($-$) the substrate gradient, and $v$ is the average run speed. 
Note that for symmetric bias, $\alpha^+ =\alpha^- = \alpha$, $\mu = v^2/\alpha d \equiv \mu_0$ in $d$ dimensions. Extension to the asymmetric case for $d>1$ is cumbersome and here we formally work only in $d = 1$. (By writing the final results in terms of $\mu_0$ the correct $d$-dependence is, however, recovered in the symmetric limit).

We connect the above expressions to the chemotactic response by modifying previous work \citep{DeGennes04, oc07} to account for the effect of agar. A bacterial run is an inhomogeneous Poisson process with rate
\be\label{chemomemo0}
\alpha_t(t) = \alpha_0 \left[1-\int_{-\infty}^tdt' K(t-t') f_\chi(x(t'))\right]\equiv\alpha_0 [1-\Delta(t)]
\ee
where the subscript $t$ indicates tumbles and, as in the main text, $\alpha_0$ is the tumble rate in the absence of bias and $f_\chi$ is a function related to substrate concentration at position $x$ via $f_\chi = s(x)/(s(x)+k_\chi)$. The function $K(t)$ is the bilobed chemotactic response function which has been measured for \colisp \citep{SegallBlockBerg82}, and obeys $\int_0^\infty K(t)dt = 0$.  The linear expression above is valid for shallow substrate gradients, i.e. the bias $\abs{\Delta(t)}\ll1$. Considering a run starting at $t=0$, in the absence of agar the probability density for a tumble occurring in the interval $[t,t+dt]$ is given by $\alpha_t(t) \exp\left(-\int_0^t  dt' \alpha_t(t')\right)$. We argue that since bacterial collisions with the agar can also be considered a Poisson process, the same probability density describes the occurrence of tumbles in agar if the tumble rate $\alpha_{t}$ is replaced by an effective rate:
\be\label{AgarTumble}
\alpha_{e}(t; C)=\alpha_{t}+\alpha_A
\ee
which comprises (independent) contributions from $\alpha_t=\alpha_t(t; C)$, the tumble rate due to the intrinsic bacterial dynamics (modulated by any chemotactic response) and $\alpha_A=\alpha_A(C)$ an additional collision rate with the agar (which also randomises swimming direction). Then the mean run duration for bacteria in agar (or other porous media) is given by
\be\label{meantime}
T(C)=\left\langle \int_0^\infty dt\,t\,\alpha_{e}(t; C) \exp\left(-\int_0^t  dt' \alpha_e(t'; C)\right) \right\rangle_{paths}
\ee
where $\langle \ldots \rangle_{paths}$ denotes an average over all possible bacterial swimming paths (the suffix will hereafter be assumed), since the nonlocal contribution $\alpha_{t}$ to $\alpha_e$ is path dependent. Then, changing variables in the memory integral \eq{chemomemo0} by defining $u=t-t'$, substituting \eq{AgarTumble} into \eq{meantime} and recalling $\abs{\Delta(t)}\ll1$, we have:
\be\label{meantime2}
T(C)\approx \frac{1}{\alpha(C)} + \alpha_0 \int_0^\infty dt \expe{-\alpha(C) t} \left\langle \int_0^t dt' \int_{0}^{\infty}du K(u) f_\chi(x(t'-u))  \right\rangle
\ee
where we have defined the unbiased tumble rate in agar
\be\label{agaralpha}
\alpha(C)=\alpha_0+\alpha_A(C).
\ee
The concentration function $f_\chi$ is related to imposed gradients by a Taylor expansion:
\be\label{f_expand}
f_\chi(x(t-u)) \approx x(t-u)  \nabla f_\chi + const.
\ee
Recalling that that $K(u)$ integrates to zero, the constant term does not contribute to integral \eq{meantime2}. Thus, following a trick introduced by De-Gennes \citep{DeGennes04}, we consider a `single delay' response function of the form $K(u)=A\delta(u-\theta)$, so \eq{meantime2} becomes
\be\label{meantime3}
T(C)\approx \frac{1}{\alpha(C)}  + A \nabla f_\chi \alpha_0 \int_0^\infty dt \expe{-\alpha(C) t} \left \langle \int_0^t dt' x(t'-\theta) \right\rangle.
\ee
Next, again following \citep{DeGennes04} (ignoring persistence and rotational diffusion, see \citep{Locsei07}) we notice that for times $t-\theta<0$ before the start of a run, the position $x(t-\theta)$ is on average not correlated to the bacterial velocity along the run. On the other hand for $t-\theta>0$, we can write $x(t-\theta)=\pm v (t-\theta)$, where $\pm v$ is the run speed up or down a gradient. \eq{meantime3} then becomes
\be\label{meantime4}
T^{\pm}(C)\approx \frac{1}{\alpha(C)}  \pm v \abs{\nabla f_\chi} \alpha_0 A \int_\theta^\infty dt \expe{-\alpha(C) t} \frac{1}{2}(t-\theta)^2 \ee
and, after integrating by parts
\be\label{meantime5}
T^{\pm}(C)\approx \frac{1}{\alpha(C)}  \pm  v\abs{\nabla f_\chi} \frac{ \alpha_0 }{ \alpha(C)^3}A \expe{-\alpha(C) \theta}  \ee.
So finally, for a general a distribution of delay times $K(\theta)$ we have
\be\label{meantime6}
T^{\pm}(C)\approx \frac{1}{\alpha(C)}  \pm  v \abs{\nabla f_\chi} \frac{ \alpha_0 }{ \alpha(C)^3}\int_0^\infty d\theta K(\theta) \expe{-\alpha(C) \theta}.  
\ee
Now we identify $\alpha^{\pm}=1/T^{\pm}$, so that we can use \eq{meantime6} and \eq{TranspMicro} to find, to leading order in $\abs{\nabla f_\chi}$,
\be\label{TranspMicroCv}
\mu(C)=\frac{v^2}{\alpha(C)};\,\,\,\,\,\,\,\, 
v_\chi(C)= v^2 \frac{ \alpha_0 }{ \alpha(C)^2} \abs{\nabla f_\chi} \int_0^{\infty}d\theta K(\theta) \expe{-\alpha(C) \theta}. 
\ee
Or, since the chemotactic sensitivity parameter $\chi$ is defined by $v_\chi=\chi(C)\nabla f_\chi$
\be\label{TranspMicroC}
\mu(C)=\frac{v^2}{\alpha_0}\left[1+\frac{\alpha_A(C)}{\alpha_0}\right]^{-1};\,\,\,\,  
\chi(C)=\frac{v^2}{\alpha_0} \left[1+\frac{\alpha_A(C)}{\alpha_0}\right]^{-2}\int_0^{\infty}d\theta K(\theta) \expe{-\alpha(C) \theta}. 
\ee
where we have expanded the agar tumble rate defined in \eq{agaralpha}. Equations \eq{TranspMicroC} are the bacterial transport parameters in agar accounting for a chemotactic response nonlocal in time. In the absence of agar ($C\to0$), the experimentally measured values of the bacterial transport parameters are $\mu_0$ and $\chi_0$, the {\it in liquido} diffusivity and chemotactic parameter. In this limit \eq{TranspMicroC} become the expressions derived by de Gennes \citep{DeGennes04}
\be\label{TranspMicro-muLim}
\mu(C\to0)=\frac{v^2}{\alpha_0}\equiv \mu_0;\,\,\,\,\,\,\,\, \chi(C\to0)= \frac{ v^2 }{ \alpha_0} \int_0^{\infty}d\theta K(\theta) \expe{-\alpha_0 \theta}  \equiv \chi_0.
\ee
Using \eq{TranspMicro-muLim}, we can rewrite \eq{TranspMicroC} as
\be\label{TranspMicroCexp}
\mu(C)=\mu_0 \left[1+\frac{\alpha_A(C)}{\alpha_0}\right]^{-1};\,\,\,\,\,\,\,\,  
\chi(C)= \chi_0 \left[1+\frac{\alpha_A(C)}{\alpha_0}\right]^{-2}I_\chi \left(\frac{\alpha_A(C)}{\alpha_0}\right),
\ee
where 
\be\label{chemointegral}
I_\chi \left(\frac{\alpha_A(C)}{\alpha_0}\right)=\frac{\int_0^{\infty}d\theta K(\theta) \expe{-\alpha_0 \left[1+\frac{\alpha_A(C)}{\alpha_0}\right]\theta} }{\int_0^{\infty}d\theta K(\theta) \expe{-\alpha_0 \theta}}.
\ee
To solve the model presented in the main text, we require an explicit expression of $K(t)$ to evaluate \eq{chemointegral}, and thus \eq{TranspMicroCexp}. We use a recently proposed fit to the experimentally measured impulse response of \colisp \citep{ClarkGrant05}, and write: 
\be\label{chemoresponsefunction}
K(t) =N_0 \expe{-\alpha_0 t} \left[1-A_0\left(\alpha_0 t + \frac{1}{2} \alpha_0^2 t^2 \right)\right],
\ee
where $\alpha_0$ is the base tumble rate, $A_0$ is a dimensionless constant and $N_0>0$ is a normalisation constant whose value is unimportant, as it cancels out in the expression for $I_\chi$. 

To facilitate the discussion of our results, we also evaluate two limiting expressions for the concentration dependence of the chemotactic parameter in \eq{TranspMicroCexp}. For very low concentrations bacterial collisions with the agar are rare, $\alpha_A(C)\ll1$ ($\alpha(C) \approx \alpha_0$), so, expanding to first order, \eq{chemointegral} becomes $I_\chi\approx 1- \kappa\, \alpha_A/\alpha_0 $, where $\kappa\equiv \int_0^{\infty}d\theta K(\theta) \expe{-\alpha_0 \theta} \alpha_0\theta/\int_0^{\infty}d\theta K(\theta) \expe{-\alpha_0 \theta}$. For large agar concentrations, on the other hand, collisions with the agar are frequent and confuse the chemotactic response. The effective tumble rate is so large compared to the natural one, $\alpha(C) \gg \alpha_0$, that $K(\theta)$ can be approximated by $K(0)$ in the numerator of \eq{chemointegral}, where the integrand falls rapidly
to zero for $\theta \ge 1/\alpha(C)$. In this case $I_\chi(\alpha_A/\alpha_0\gg1)\approx \beta [1+\alpha_A(C)/\alpha_0]^{-1}$, where $\beta=K(0)/(\alpha_0 \int_0^{\infty}d\theta K(\theta) \expe{-\alpha_0 \theta})$. We can then write asymptotic expressions for the chemotactic parameter:
\bea\label{TranspMicro-Lim1}
\chi(C)\simeq
\begin{cases} 
\chi_0 \left[1+\frac{\alpha_A(C)}{\alpha_0}\right]^{-2}\left[1- \kappa \frac{\alpha_A(C)}{\alpha_0}\right] & \text{if $\alpha(C)\approx \alpha_0$}\\\label{TranspMicro-vLim1}
\beta\, \chi_0 \left[1+\frac{\alpha_A(C)}{\alpha_0}\right]^{-3}&\text{if $\alpha(C)\gg \alpha_0$ }.
\end{cases}
\eea
If, as in the main text, the values $A_0=0.5$ and $\alpha_0=1$ are used in \eq{chemoresponsefunction}, then $\kappa=1/10$ and $\beta=16/5$. 

\end{appendix}


\clearpage
\section*{Figure Legends}
\subsubsection*{Figure~\ref{015-04radprof}.}
(a)  Early (top row) and advanced (bottom) stages of the migration of {\it E. coli} AB1157 populations through LB agar of concentration $C=0.15$-$0.35\%$, as labelled. Shown are circular views ($65$ mm diameter) from minimally processed images (see methods) of $100$ mm diameter petris filled with $10$ mm thick agar. (b-e) Azimuthally averaged radial intensity profiles from the images (see text). The time since inoculation in hours is indicated throughout.

\subsubsection*{Figure~\ref{MorphoHighC}.}
Bacterial populations for $C = 0.4$-$0.5\%$. Colonies (65~mm views) are initially circular (top row), but quickly develop blebbing instabilities (bottom row). Images were minimally processed as for \fig{015-04radprof}a (see methods).

\subsubsection*{Figure~\ref{Rvst015-04}.}
Colony radius, $r$, against the time, $t$, since growth inception (see text) for $C=0.15$-$0.4\%$, as shown. The inset shows a linear fit to the raw radial data for $C=0.3\%$ against time since inoculation, $t_i$. Similar fits for all other concentrations define the migration speed (slope) and the latency time $\Delta t_{l}$ (intersection with the time axis). Error bars are at most the size of a data point.

\subsubsection*{Figure~\ref{VvsCa}.}
Experimental migration front speed as a function of concentration in the range $0.15$-$0.35\%$ together with a best fit to the data using our model. Also shown for best fit parameters are the model `efficient' and `confused' limits, and the prediction from gas kinetic models \citep{BartonFord96}. The inset shows the same data but including points for $C=0.4$-$0.5\%$, labelled differently to indicate a different mode of migration at these concentrations. The model breakdown in this region is evident.

\begin{figure}[tbph] 
\begin{center}  
\includegraphics[width=0.327\linewidth]{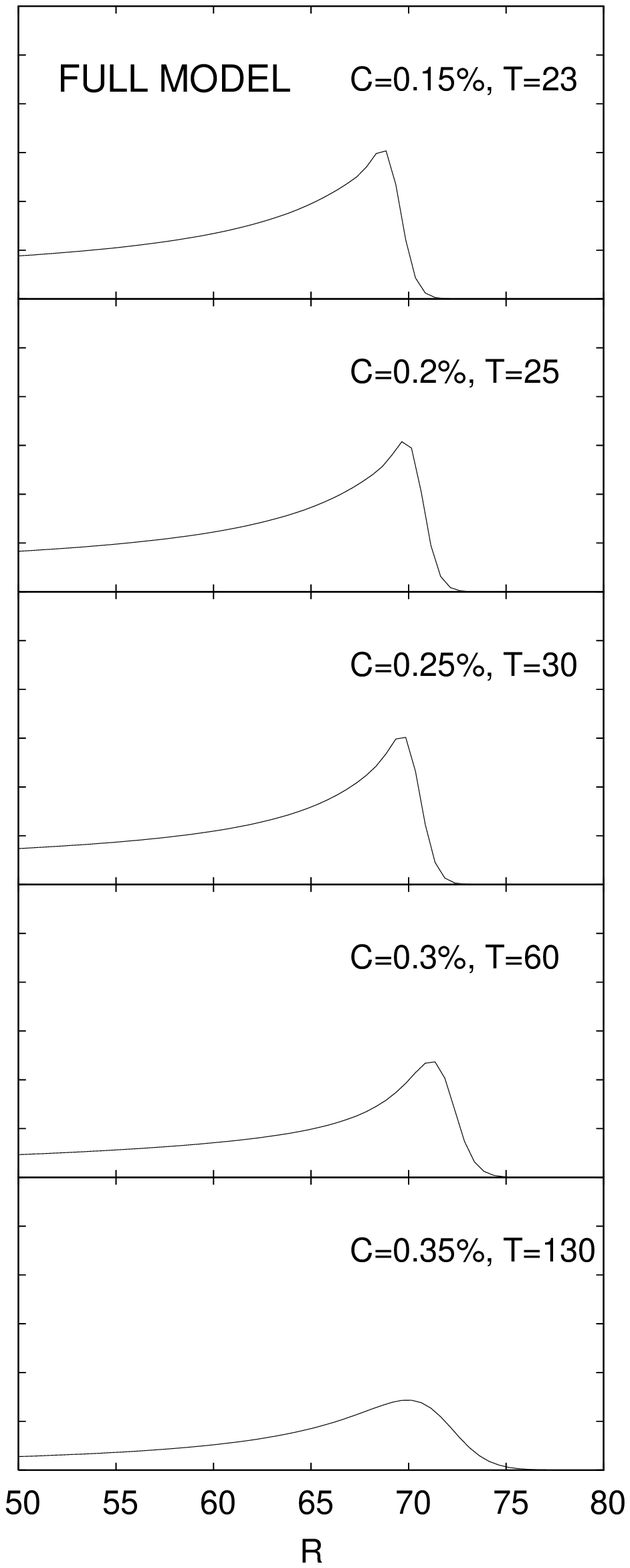}
\includegraphics[width=0.327\linewidth]{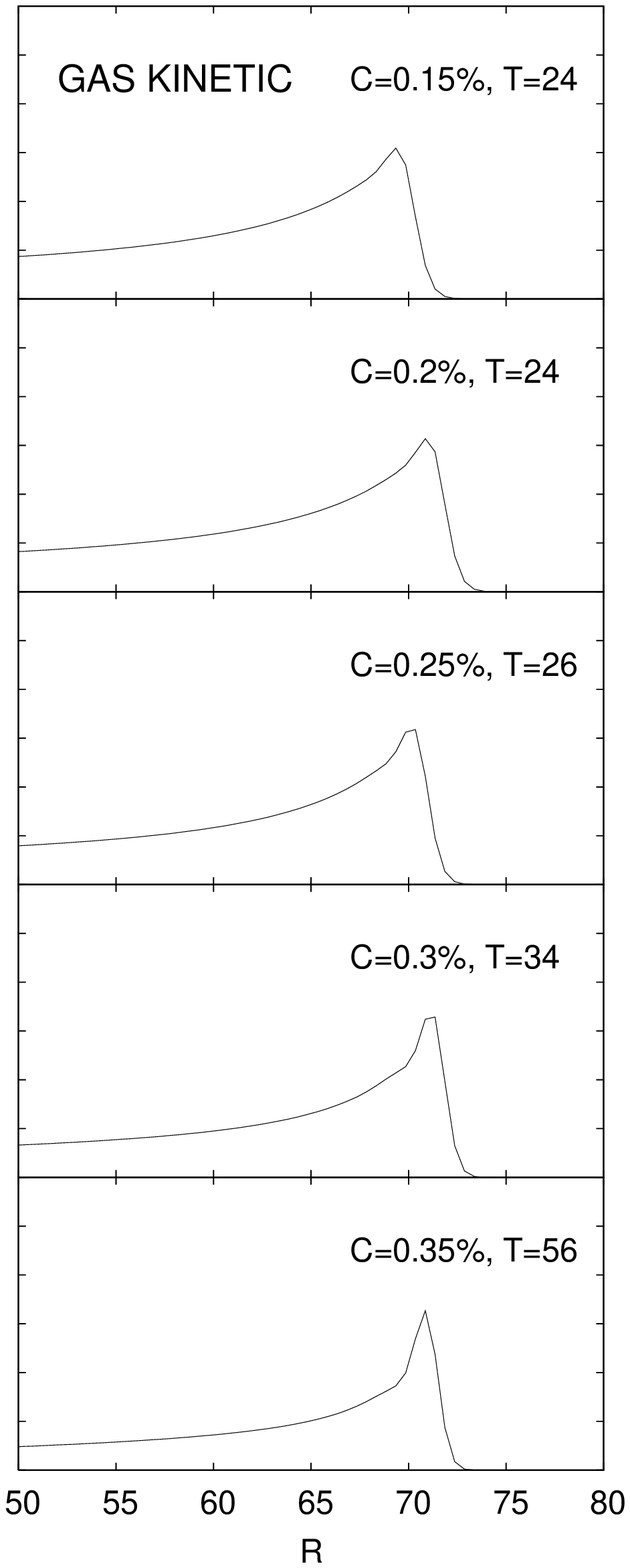}
\caption{}
\label{thprofiles} 
\end{center}
\end{figure}

\subsubsection*{Figure~\ref{thprofiles}.}
\begin{figure}
\begin{center}
\includegraphics[width=\linewidth]{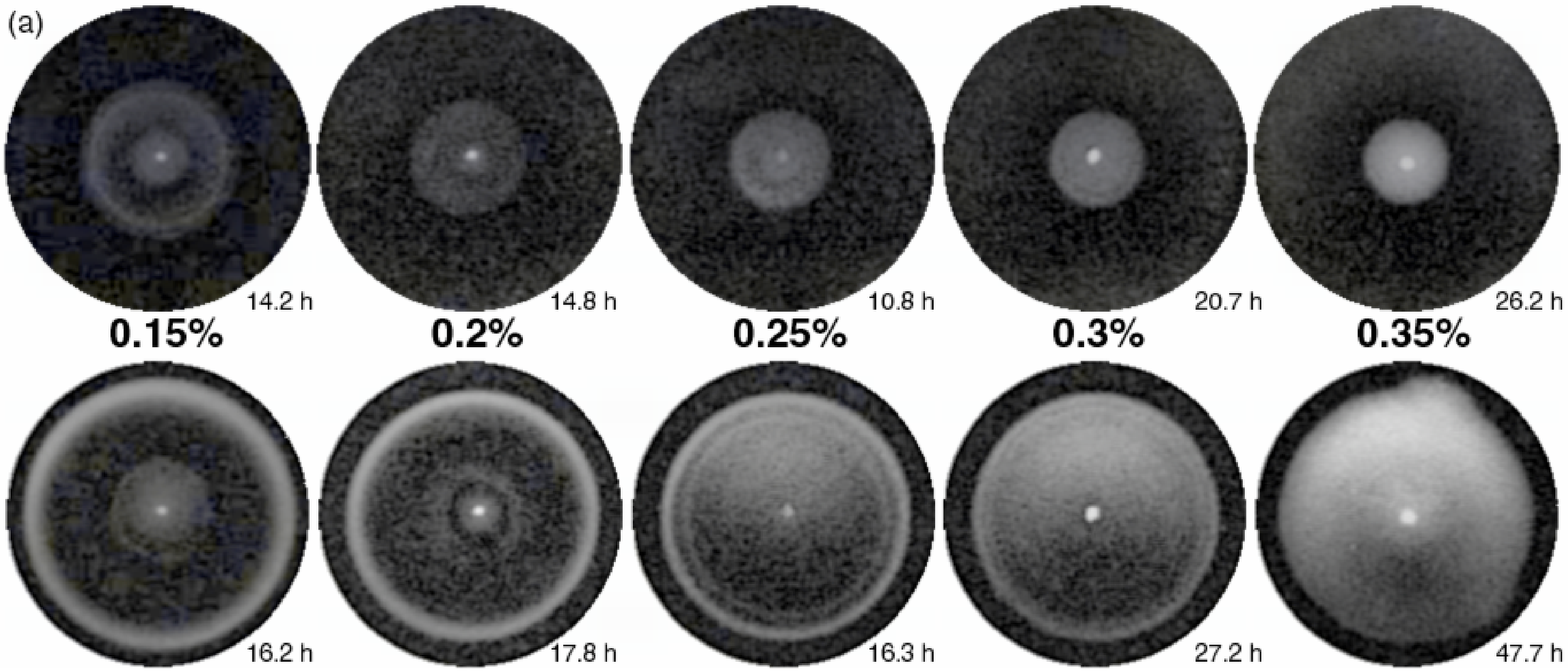}\\
\includegraphics[width=0.327\linewidth]{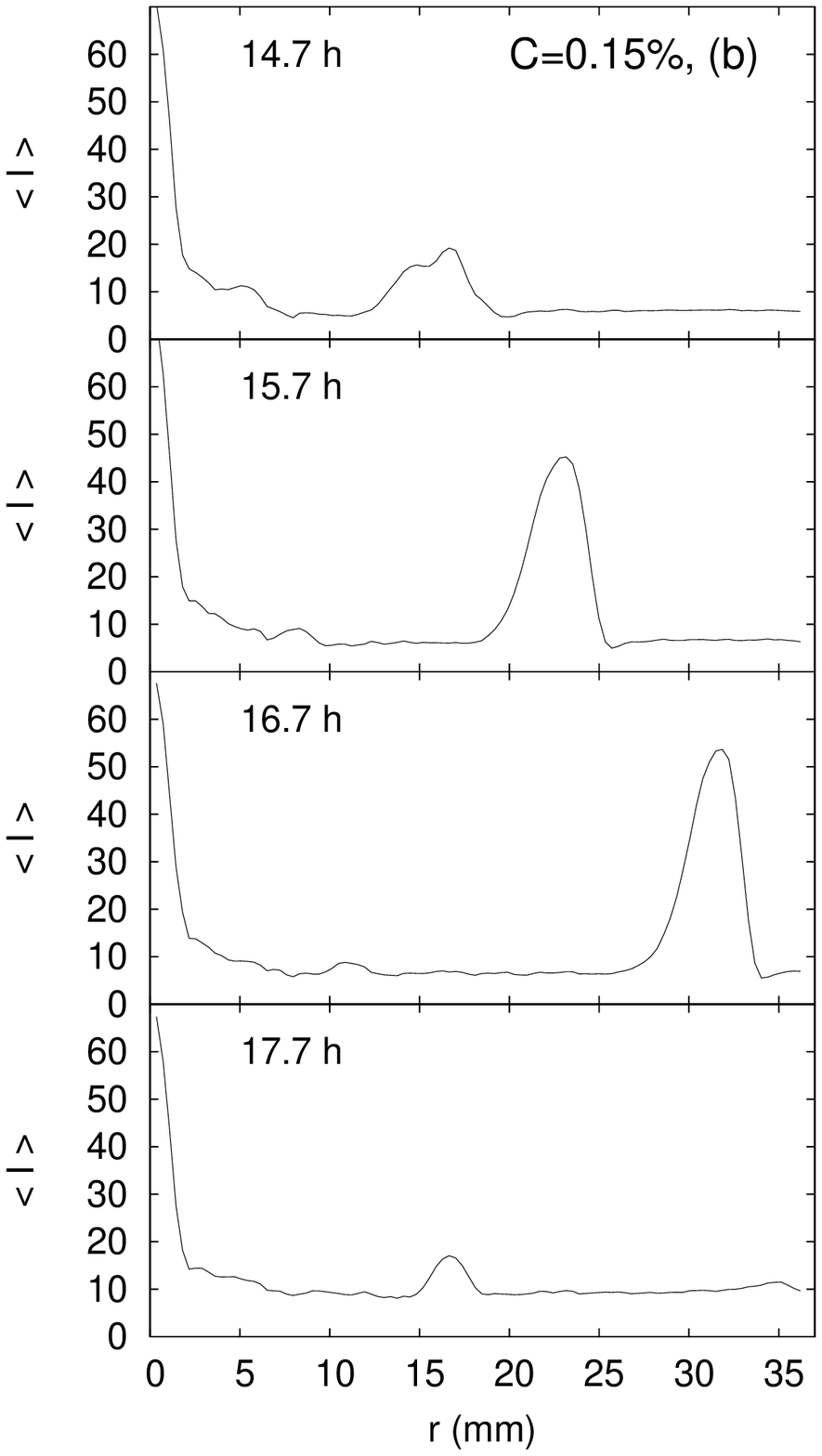}
\includegraphics[width=0.327\linewidth]{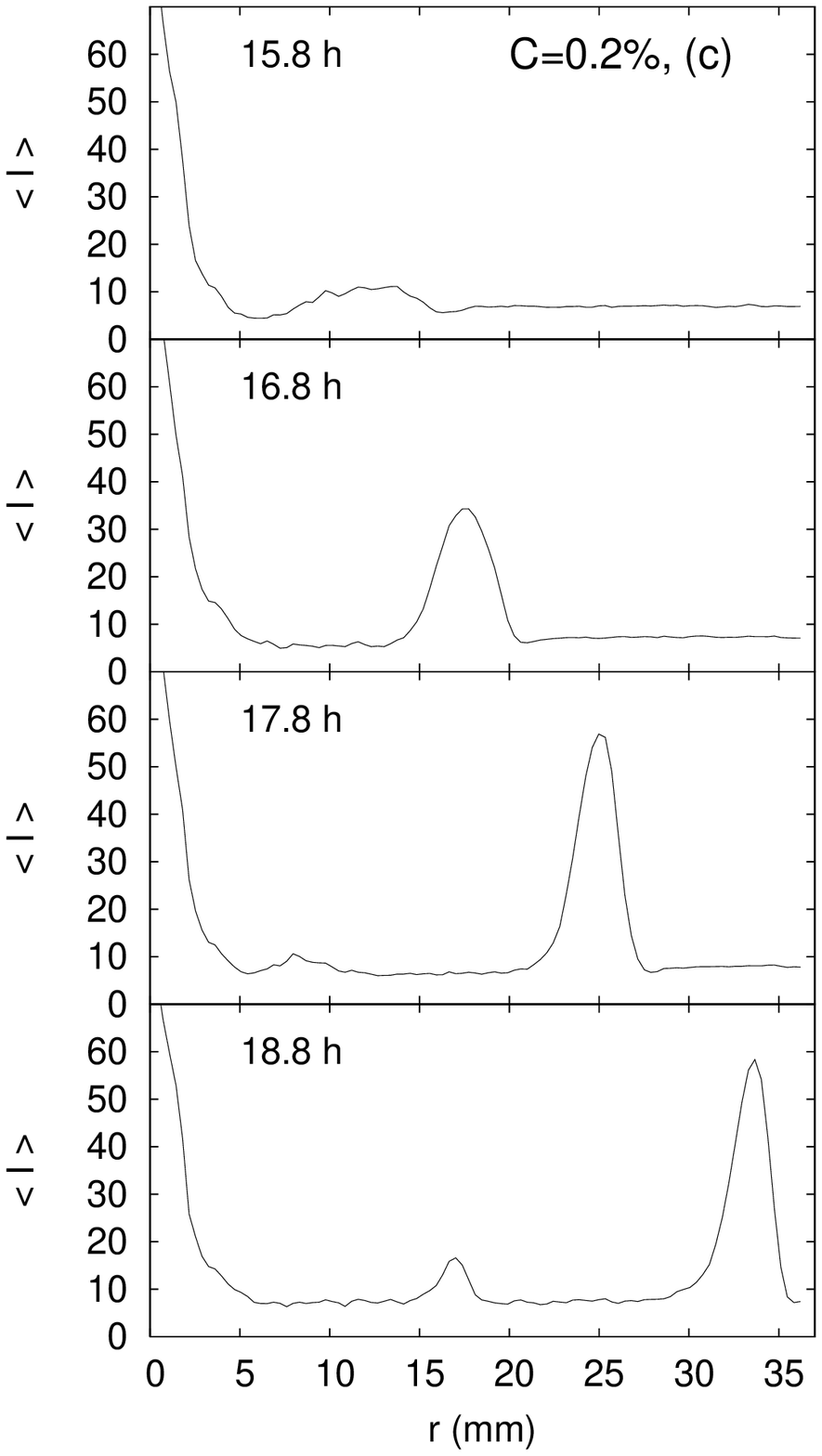}
\includegraphics[width=0.327\linewidth]{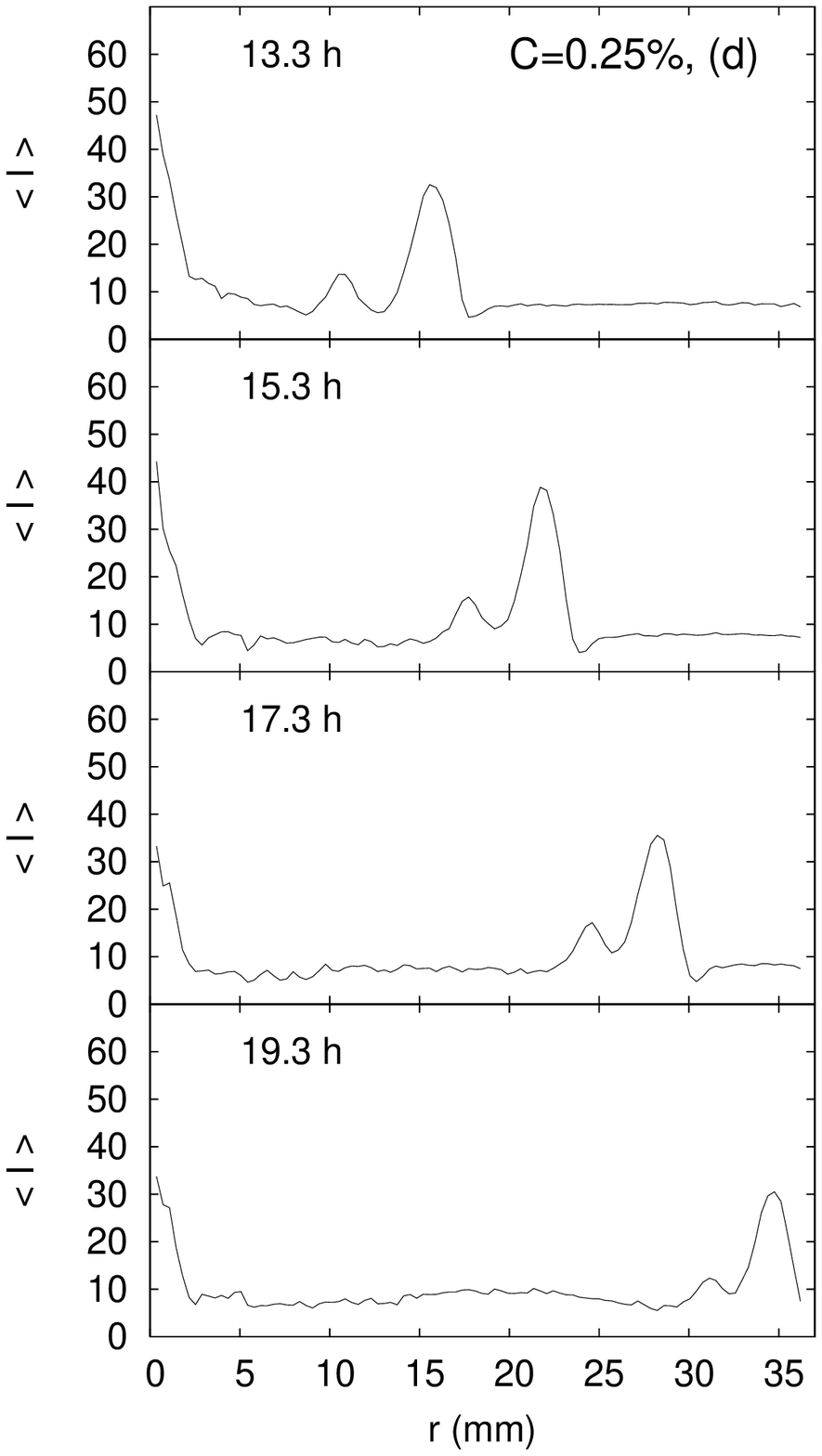}\\
\includegraphics[width=0.327\linewidth]{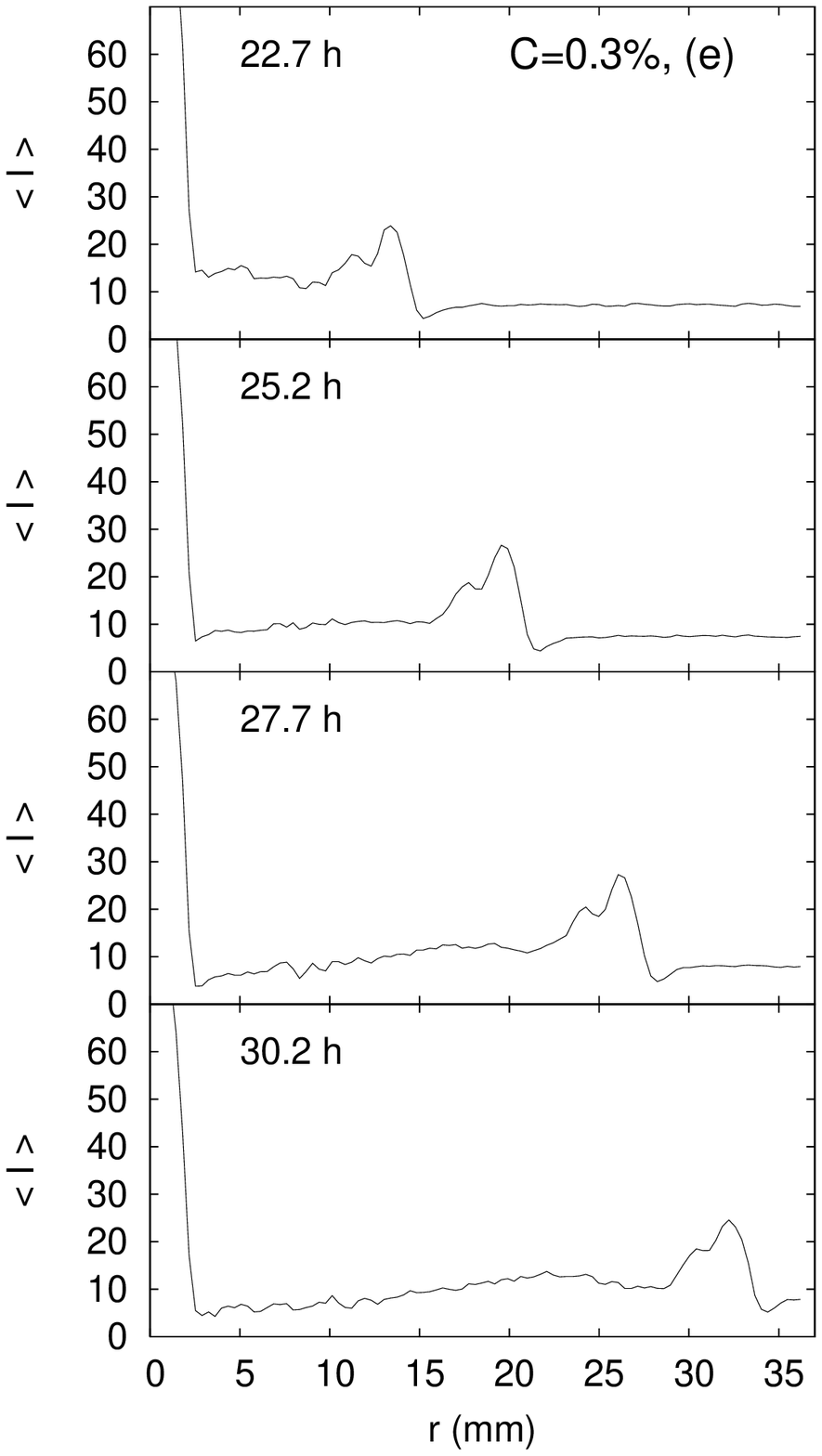}
\includegraphics[width=0.327\linewidth]{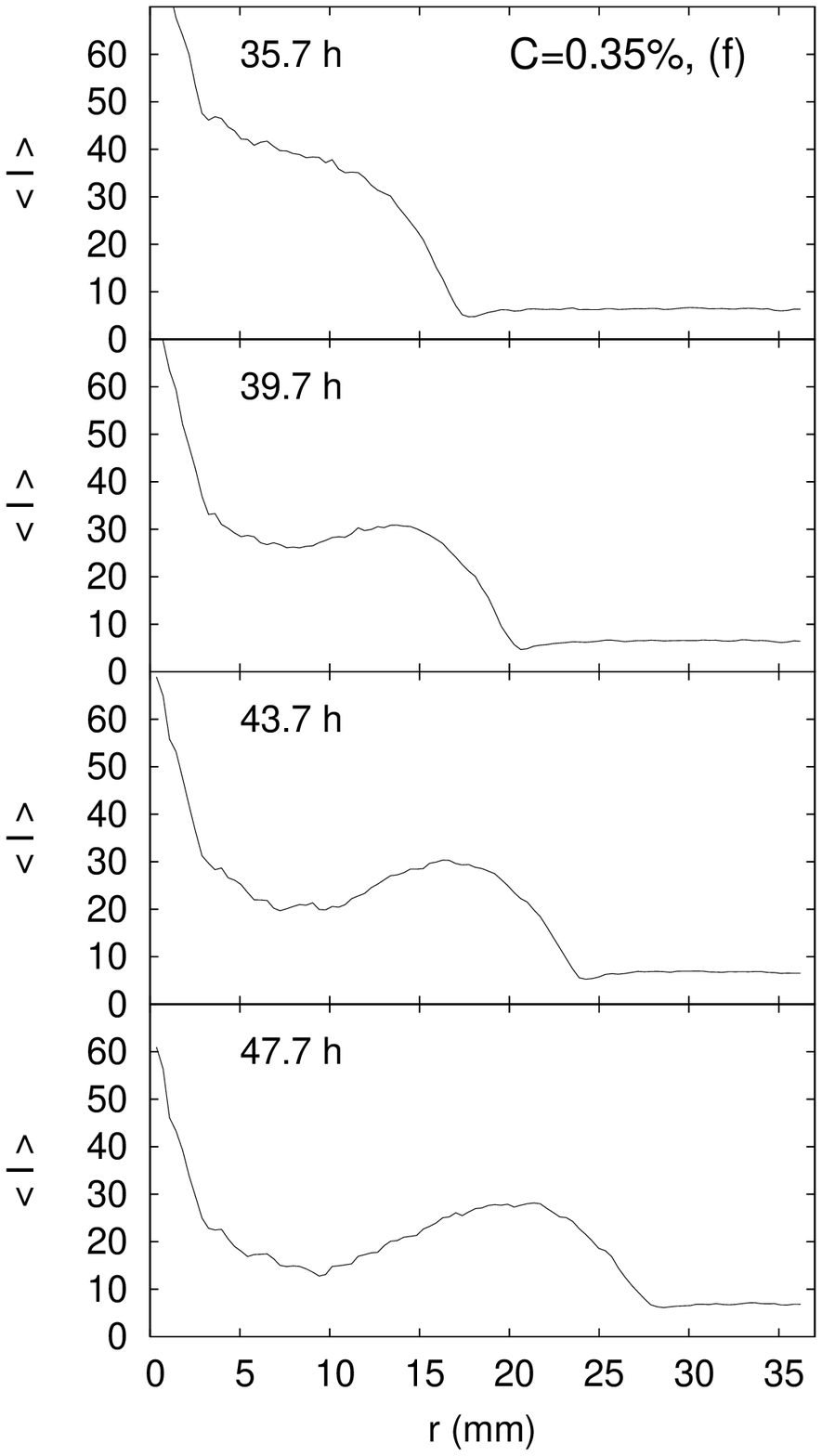}
\caption{}
\label{015-04radprof} 
\end{center}
\end{figure}

Theoretical predictions for the band profiles for the full model (left) and gas kinetic models \citep{BartonFord96} (right) in the same range probed in experiments: $C = 0.15$-$0.35\%$, as indicated. In the full model, as concentration is increased the dynamics changes from chemotactic (sharp bands) to growth/diffusion dominated (broad bands). This gradual transition is qualitatively the same as observed in experiment (see \fig{015-04radprof}), and is not predicted by the gas kinetic model.

\clearpage

\clearpage

\clearpage

\clearpage

\clearpage

\end{document}